# Measurement of the Speed of Gravity

Yin Zhu

*Agriculture Department of Hubei Province, Wuhan, China*

waterzhu@163.com

**Abstract:** The speed of gravity is an important universal constant. But, it has not been directly known with experiment or observation. The explanations for it are contradicted with each other. Here, it is presented that the interaction and propagation of the gravitational field could be tested and understood by comparing the measured speed of gravitational force with the measured speed of Coulomb force. A design to measure the speeds of gravitational and Coulomb force is presented. From satellite motions, it is observed that the speed of gravitational force is larger than the speed of light in a vacuum. From this observation and the recent experiments, the structure of electric and gravitational fields is studied. A line to indirectly test the wavelengths of gravitational waves is presented.

Fomalont and Kopeikin[1] in 2002 claimed that to 20% accuracy they confirmed that the speed of gravity is equal to the speed of light in vacuum. Their work was immediately contradicted by Will[2] and other several physicists.[3-7] Fomalont and Kopeikin[1] accepted that their measurement is not sufficiently accurate to detect terms of order $v^2/c^2$, which can experimentally distinguish Kopeikin interpretation from Will interpretation. Fomalont et al[8] reported their measurements in 2009 and claimed that these measurements are more accurate than the 2002 VLBA experiment [1], but did not claim that the terms of order $v^2/c^2$ have been detected.

Within the post-Newtonian framework, several metric theories have studied the radiation and propagation of gravitational waves.[9] For example, in the Rosen bi-metric theory,[10] the difference between the speed of gravity and the speed of light could be tested by comparing the arrival times of a gravitational wave and an electromagnetic wave from the same event: a supernova. Hulse and Taylor [11] showed the indirect evidence for gravitational radiation. However, the gravitational waves themselves have not yet been detected directly.[12] In electrodynamics the speed of electromagnetic waves appears in Maxwell equations as $c = \sqrt{\mu_0 \varepsilon_0}$, no such constant appears in any theory of gravity. Currently, our only experimental insights into the speed of the gravitational waves are derived from observations of the binary pulsar PSR 1913+16. Measurements of its orbital decay agree at the ~1% level with general relativity (GR), assuming that energy is emitted in the form of quadrupole gravitational radiation.[13,14]

At the low speeds and in the weak-field limit, the Newtonian limit can be derived from Einstein field equations.[15] In this letter, it is shown that, the speed of gravity can be studied or measured with the retarded gravitational potential.

In the Newtonian theory of gravity, the gravitational interaction is instantaneous. Attempting to combine a finite gravitational speed with Newton's theory, Laplace[16] concluded that the speed of gravity is $c_g \geq 7 \times 10^6 c$ by analyzing the motion of the Moon in 1805. Lightman[17] et al determined that with the purely central force, if the speed of gravity is finite, the forces in the two-body system no longer point toward the center of mass, and make orbit unstable. In 1998, Flandern[18] set a limit $c_g \geq 2 \times 10^{10} c$ for the speed of gravity, based on a purely central force of Newtonian theory from the observations of the solar system and binary stars. However, in Einstein's general relativity the speed of gravity is postulated to be equal to the speed of light in vacuum.[19] Compared with the electromagnetic field, Carlip,[20] Marsh and Nissim-Sabat,[21] and Ibison et al[22] argued that in GR the gravitational interaction propagates at the speed of light, but the velocity-dependent terms in the interaction cancel the aberration.

Comparing the gravitational field with the electromagnetic field is a current way to understand and to study the speed of gravity.[1, 10, 15, 20-22] Here, I show that by comparing the measured speed of the Newtonian gravitational force with the measured speed of Coulomb force, some evidences for the speed of propagation of the gravitational and electromagnetic field (waves) can be found.

In GR, in the weak field limit, for the slowly moving mass, the gravitational field can be expressed as a perturbation of the space-time metric. Adopted the units c=1, which can be written as

$$g_{\mu\nu} = \eta_{\mu\nu} + h_{\mu\nu}, \qquad |h_{\mu\nu}| \ll 1, \tag{1}$$

where $\eta_{\mu\nu}$ is the Minkowski metric, $\eta_{\mu\nu} = diag(-1, +1, +1, +1)$, and $h_{\mu\nu}$ is the perturbation. Einstein's field coupled to matter is

$$\bar{h}_{\mu\nu} = -16\pi G T_{\mu\nu}, \tag{2}$$

Where $\bar{h}_{\mu\nu}$ is the "trace-reversed" perturbation, $\bar{h}_{\mu\nu} = h_{\mu\nu} - \frac{1}{2}\eta_{\mu\nu}h$, and $T_{\mu\nu}$ is the stress-energy tensor.

Solving Eq. (2) by defining the quadrupole moment tensor of the energy density of the source as

$$q_{ij}(t) = \int x^i x^j T^{00}(t,x) d^3 x.$$

In terms of the Fourier transform of the quadrupole moment, and transform back to t, the solution to Eq. (2) takes on the compact form

$$\bar{h}_{\mu\nu}(t,x) = \frac{2G}{R}\frac{d^2 q_{ij}}{dt^2}(t_m), \tag{3}$$

where M is the mass of the moving source, R is the distance between the source and the observer, G is the gravitational constant; and

$$t_m = t - R, \tag{4}$$

where $t_m$ is the time that the source mass is located at the retarded position, t is the time that the observer is affected by the corresponding potential.

From Eq. (1), in a vacuum, we can recover the conventional relativistic wave equation

$$\Box h_{\mu\nu} = 0, \tag{5}$$

where $\Box$ is the d'Alembertian operator.

In the Newtonian limit, we have

$$h_{ij} = -2\Phi\delta_{ij}, \tag{6}$$

Where $\Phi$ is the conventional Newtonian potential, $\Phi = -GM/R$, $\delta_{ij}$ is the Kronecker symbol.

The flat-space d'Alembertian operator has the form $\Box = -\frac{\partial^2}{\partial t^2} + \nabla^2$, from Eqs. (5) and (6), we have

$$\frac{\partial^2 \Phi}{\partial t^2} - \nabla^2 \Phi = 0. \tag{7}$$

Eq. (7) just is one of the d'Alembertian waves equation of the gravitational waves.

Analogy with the retarded potential of the electromagnetic field, letting the observer at the field point, and neglecting the speed of the source, from Eqs. (3), (4), (6) and (7), the retarded potential of the gravitational field can be written as [23]

$$\Phi(x,t) = -G\frac{M(t_m)}{R}, \tag{8}$$

$$t_m = t - \frac{R}{c_g}. \tag{9}$$

where $c_g$ is the speed of gravity, $c_g = c$.

This deducing shows that, in the low speeds and weak-field limit, the curvature space-time [Eqs.(1-3)], gravitational waves [Eqs.(5-7)], the speed of gravity [Eqs.(8,9)], as well as the gravitational field, can be described and studied with the Newtonian gravitational potential.

From Eqs. (8) and (9), treating the Sun (or the Moon) as effectively point mass, its Liénard-Wiechert potential can be written as [2]

$$\phi(\vec{r},t) = \frac{GM}{|\vec{r}-\vec{r}_m(t_m)| - \vec{v}\cdot(\vec{r}-\vec{r}_m(t_m))/c_g}, \tag{10}$$

where $t_m$ is "retarded" time given by the implicit equation

$$t_m = t - \frac{|\vec{r}-\vec{r}_m(t_m)|}{c_g}, \tag{11}$$

and $\quad |\vec{r} - \vec{r}_m(t_m)| = R = \sqrt{[x - x_m(t_m)]^2 + [y - y_m(t_m)]^2 + [z - z_m(t_m)]^2}, \quad \vec{v} = \frac{\partial \vec{r}_m(t_m)}{\partial t_m}.$

We can obtain Newtonian gravitational field strength from $h = \nabla \Phi(\vec{r}, t_m)$. The potential $\Phi$ is an explicit function of $\vec{r}$ and $t_m$:

$$\Phi = \Phi(\vec{r}, t_m).$$

Thus, 
$$\nabla \Phi(\vec{r}, t_m) = \nabla_{\vec{r}} \Phi(\vec{r}, t_m) + \frac{\partial \Phi}{\partial t_m} \nabla t_m,$$

$$\frac{\partial t_m}{\partial t} = \frac{\partial}{\partial t}\left(t - \frac{|\vec{r} - \vec{r}_m(t_m)|}{c_g}\right)$$

$$= 1 - \frac{1}{c_g}\frac{\partial R}{\partial t},$$

where 
$$\frac{\partial R}{\partial t} = -\frac{\partial t_m}{\partial t}\vec{n} \cdot \vec{v}, \quad \vec{n} = \frac{\vec{R}}{R},$$

Substituting $\partial R / \partial t$ into the $\partial t_m / \partial t$ expression, we obtain

$$\frac{\partial t_m}{\partial t} = \frac{1}{(1 - \vec{n}\cdot\vec{\beta})} = \frac{1}{k},$$

where 
$$k = 1 - \vec{n} \cdot \vec{\beta}, \quad \vec{\beta} = \frac{\vec{v}}{c_g}.$$

After a routine derivation, we can obtain the expression for the gravitational field: [20, 24]

$$h = GM\left\{\frac{1}{k^3\gamma^2 R^2}(\vec{n} - \vec{\beta}) + radiative\ terms\right\}, \tag{12}$$

where 
$$\gamma^2 = 1 - \beta^2, \quad \beta = \frac{v}{c_g}.$$

The Newtonian gravitational field strength $h_{str}$ is

$$h_{str} = G\frac{M}{R^2}\frac{1}{k^3\gamma^2}(\vec{n} - \vec{\beta}) \tag{13}$$

According to Carlip,[20] the terms $(\vec{n} - \vec{\beta})$ could be interpreted, i.e., the retarded direction $\vec{n}$ is linearly extrapolated toward the "instantaneous" direction; the gravitational acceleration directed toward the retarded position of the source is extrapolated toward its instantaneous position. In this case, "if a source abruptly stops moving at a point $z(s_0)$, a test particle at position x will continue to accelerate toward the extrapolated position of the source until the time it takes for a signal to propagate from $z(s_0)$ to x at light speed."[20] In other words, during a solar/lunar eclipse when the Sun, Moon and Earth are on the straight line EMS, observed at a point E on the Earth, the Sun is located at point $S(r, t)$, but, the Newtonian force of the Sun on the Earth is from that the Sun was locating at point $S'(r_m, t_m)$, and $t - t_m = R/c_g \approx 500s$, $\theta = arccos\frac{v}{c_g}$. where R is the distance between the Sun and the Earth, v is the speed of the Sun relative to the observer (which can be treated approximately as a uniform motion), $c_g = c$, $\theta$ is an angle between straight line EMS and ES'.

Thus, Eqs. (11) and (13) mean that, during a solar/lunar eclipse, the speed of the maximum Newtonian gravitational strength (force) along the line EMS is infinite. Assuming that P and E are two different points on the line EMS, from Eqs. (11) and (13) we know, the maximum Newtonian gravitational strength (force) of the Sun (or Moon) at points $P(t)$ and $E(t)$ emerge at the same time, t. The time duration $\Delta t=0$, the measured speed of this largest gravitational force along the straight line EMS is infinite.

This conclusion is consistent with the observations of the absence of the gravity aberration in the solar system, which needs that the speed of the Newtonian gravitational force is infinite (or much faster than c under the assumption of pure central force).

The gravitational force between atom and a mass with 500kg can be measured with atom interferometer. Thus, with atom interferometer, the speed of gravitational force can be measured in laboratory. Here, a detailed design is presented to measure this speed and the speed of Coulomb force.

With atom interferometer, the gravitational force can be demonstrated with a moving mass and atom as shown in Fig.1a. It intuitively shows whether the force of the moving mass acting on the atoms is equal to c or larger than c by observing the times of the motions of the atoms reflecting to the moving of the mass. And, the speed of gravitational force also can be measured with the design as shown in Fig.1b. The interaction of the electrostatic fields can be demonstrated with the design shown in Fig.1b as the atoms and mass are replaced by the charged balls. The speed of Coulomb force can be measured with this demonstration.

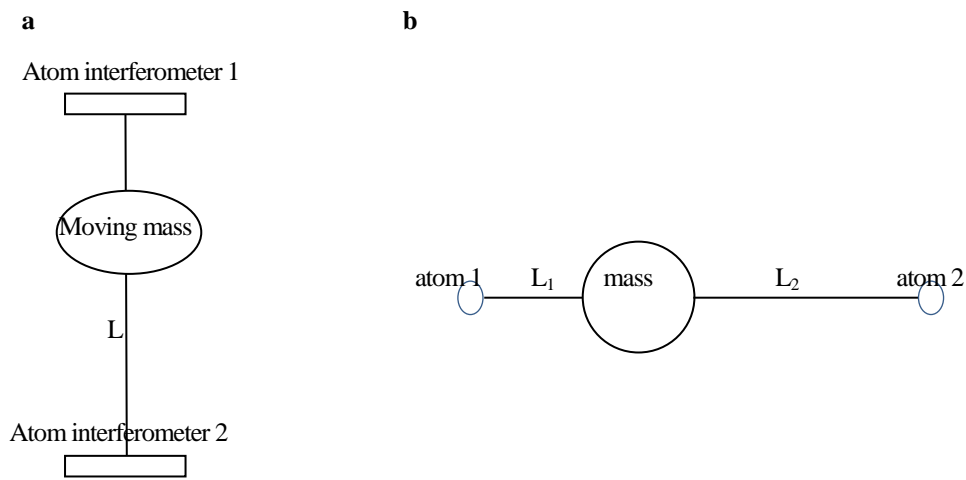

**Fig.1**. **a**. **Scheme of the experiment.** L is a distance between atom interferometers 1 and 2. L is vertical to the surface of the Earth. A mass is moving up and down along L. The motions of the atoms in the interferometers can be affected by the moving of the mass. This design can be level positioned and the mass can be moved along the direction vertical to L. **b.** The mass and atoms 1 and 2 are stationary and they are gravitating each other. Once the mass is moving, the atoms 1 and 2 shall be moved at time $t_1=L_1/c_g$ and $t_2=L_2/c_g$ respectively.

The speed of gravitational force can be measured from $c_g=(L_2-L_1)/(t_2-t_1)$. For measuring the speed of Coulomb force, we can replace the atoms 1 and 2, the mass by charged balls 1, 3 and 2. Ball 2 can move along the direction vertical to L. Balls 1 and 3 are the keys controlling two lines of interfering light.

The gravitational force between the mass M and the atom m is $F=GMm/R^2$. (it can be written as the formula for the distribution of the mass as many atoms are interacting with the mass[25]) R is varied with the mass moving up and down, which results in that the force F measured within the interferometers is varied periodically as shown in Fig.2. The lines 1 and 2 in Fig. 2 are measured with atom interferometers 1 and 2 respectively. $t_1$ is relative to the largest gravitational force of the moving mass acting on the atoms in line 1 while $t_2$ relative to the least one in line 2. If the gravitational force is instantaneous, the times of the motions of the atoms in interferometers 1 and 2 affected by the moving mass are same, $t_2-t_1=0$. On the other hand, if this force is finite, there is $t_2-t_1 \neq 0$ for that. Besides the mass is at the middle of L, the variations of the distance between the moving mass and the atoms is $dL \neq 0$. The speed of the gravitational force can be measured from $c_g=L/(t_2-t_1)$ as the mass is moving at one of the ends of L. A series of $t_{1i}$ and $t_{2i}$ relative to the largest or least gravitational force from the moving mass can be measured as the mass is continually moving up and down, where $i=1 \rightarrow n$. As $t_{1i}$ is relative to the largest force, $t_{2i}$ is relative the least force, $t_{2i}-t_{1i}>0$, and vice versa. A more precision value of the speed of gravitational force can be obtained from the average value of $(t_{2i}-t_{1i})$.

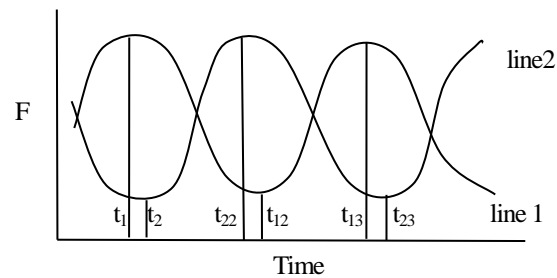

**Fig.2**. The lines of measured data. The lines 1 and 2 are measured respectively with interferometers 1 and 2. The mass is moving up and down. As the distance between the mass and interferometer 1 is the shortest, the force of the mass acting on the atoms in interferometer 1 is the largest while that in interferometer 2 is the least, and vice versa. $t_1$ is relative to the largest force in line 1 while $t_2$ relative to the least one in line 2. If the speed of Newtonian gravitational force is instantaneous, $t_2-t_1=0$. If this speed is finite, $t_2-t_1>0$. A series of data of the times $t_{1i}$ and $t_{2i}$ relative to the largest and least gravitational force from the continual moving of the mass up and down.

For precisely measuring the Newtonian gravitational universal constant and gravitational acceleration with atom interferometer, several experiments[25-31] within the design like that shown in Fig. 1a were performed. In these experiments, the variation of time at the level of $4.17 \times 10^{-11}$ seconds relative to the variation of the gravitational force has been measured, and it can be measured at the level of $\sim 10^{-15}$ seconds by comparing with a reference. (Please see the Appendix 1) And, measured with the atom interferometers 1 and 2, the motion of the atom reflecting to the mass with almost 500kg is measured. Typically, in these experiments, L is $3\times10^{-1}$m. Therefore, a speed of gravitational force at the level of $c_g=L/(t_2-t_1)=10^3c$ can be easily measured as the measured time duration is at the level of $\sim10^{-12}$ seconds with the design as shown in Fig.1. (The time duration at the level of $\sim10^{-12}$ seconds can be easily measured under the condition that it can be measured at the level of $\sim10^{-15}$ or $10^{-18}$ seconds. please see the Appendix)

The speed of gravitational force can be measured from another way. The direction of the largest gravitational force of the Sun and Moon acting on the Earth during a solar eclipse along the straight line Sun-Moon-Earth is certain and the tidal force is always with a certain direction. In this case, two atom interferometers can be positioned on different height above the surface of the Earth, the variations of time relative to the variations of gravitational force of the largest gravitational or tidal force at different height can be measured with the atom interferometers as shown in Fig.3a. In substance, the design in Fig.3a is just as that shown in Fig.1a as the moving mass is replaced by the Sun and Moon and two interferometers are positioned on one side.

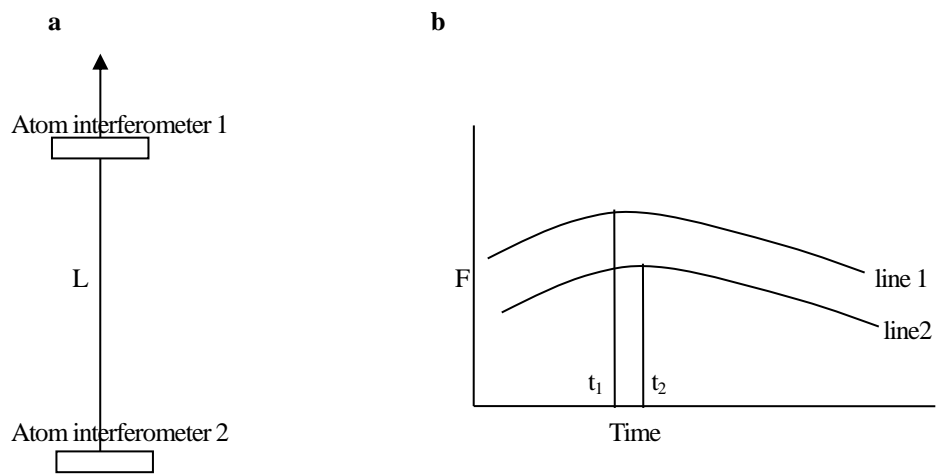

**Fig.3**a. **Scheme of measuring the speed of gravitational force during solar eclipse.** L is a distance between atom interferometers 1 and 2. Line L is vertical to the surface of the Earth. The times $t_1$ and $t_2$ of largest gravitational force of the Sun and Moon acting on the atom during a solar eclipse can be measured by atom interferometers 1 and 2 respectively. The times relative to other characteristic values of gravitational (tidal) forces also can be used to measure this speed. For example, the tidal force varies with a periodic line and the largest and least tidal force can be measured periodically. **b. The lines of measured data.** The shapes of two lines are same. $t_1$ and $t_2$ are relative to the largest gravitational force of lines 1 and 2 respectively, the gravitational force measured in line 1 is larger than that in line 2 . The shape of the lines depends on the measured object. A symmetric line can be measured during that the Sun, Moon and Earth are on a straight line and the tidal force is vertical to the surface of the Earth, and an un-symmetric line is correspondent to other status. The shape of the line is helpful for analyzing the measured data.

The force of the Sun and Moon acting on the atom is $F=G(M_{sun}+M_{moon})m/R^2$. The lines 1 and 2 in Fig. 3b are measured with atom interferometers 1 and 2 respectively. Just as shown in Fig.2, the lines also intuitively show whether the speed of gravitational force is c or is instantaneous. If this speed is instantaneous, $t_2-t_1=0$. If this speed is c, $t_2-t_1 \neq 0$ can be measured. On the surface of the Earth, some of buildings are higher than $6 \times 10^2$m. In the space, the spacecraft can be at the height higher than $3.6 \times 10^7$m (for instance, a geostationary satellite) above the surface of the Earth. Therefore, a speed of $2 \times 10^6$c (or $2 \times 10^{12}$c) on the surface of the Earth and a speed of $1.2 \times 10^{11}$c (or $1.2 \times 10^{17}$c) in a spacecraft can be measured for the gravitational force under the condition that $t_2-t_1=1 \times 10^{-12}$ (or $1 \times 10^{-18}$) seconds can be measured.

The speed of gravitational force measured from the three statuses is accordant with each other for that the experiments[25-31] proved that the Newtonian gravitational constant and acceleration measured with the design as shown in Fig. 1a is accordant with that measured on the surface of the Earth and with the observations of the motions of planets. Because a precision of $1 \times 10^{-15}$ (or $1 \times 10^{-18}$) seconds can be measured with the atom interferometer (or optical atomic clock), (Please see the Appendix) the measurable time duration listed in Table 1 can be measured easily.

For more precisely measuring the speed of gravitational force, we can let the time for the measurement long enough to make the measured lines of the variation of force relative to the time long enough. 1) Taking a lot of points on the lines, we can obtain an average value of this speed by comparing the variations of a time duration $\Delta t_1$ with that of $\Delta t_2$. 2) By comparing the measured shapes between the lines 1 and 2, or by comparing the measured shapes of the lines with the predicted shapes of the lines. 3) Taking the same long lines from lines 1 and 2 around the largest or least force to obtain the average times relative to these lines as the largest or least values are difficult to be detected.

The speed of Coulomb force can be directly measured by moving ball 2 as shown in Fig.1b. Because the distance between balls 2 and 3 is longer than that between balls 1 and 2, the time duration $\Delta t$ between the motions of ball 1 and 3 affected by the moving of ball 2 could be observed. As $L_2 > L_1$, if the speed of Coulomb force is c, $\Delta t > 0$; if this speed is infinite or much large than c, $\Delta t = 0$.

The time $\Delta t$ can be measured with light interfering. The time duration between two lines of interfering light can be controlled by balls 1 and 3 as the two balls are the keys for the two lines of interfering light respectively. A time of $1 \times 10^{-12} \sim 1 \times 10^{-13}$ seconds can be measured with light interfering.[32] The time duration from the speed of Coulomb force $c_c$ is $\Delta t = \Delta L/c_c$. Under the condition of $\Delta L = L_2 - L_1 = 3 \times 10^{-1}$m, if $c_c = c$, $\Delta t = 1 \times 10^{-9}$ seconds. It is very easy to be observed. If no fringe for the light interfering is observed, it indicates that $\Delta t < 1 \times 10^{-12}$ seconds and $c_c > 10^3$c or infinite. $\Delta t$ also can be measured with high-speed camera which can measure the time at the precision $1 \times 10^{-12}$ seconds.[32]

The above analyses show the limits for the measurable speeds of gravitational and Coulomb force as shown in Table 1.

**Table 1. The limit for measurable speeds of gravitational and Coulomb force**

|  | Speed(c) | Distance(m) | $(t_2-t_1)$ (s) |
|---|---|---|---|
| Gravitational force | $10^3$, $2 \times 10^6$, $\sim 10^{11}$ | $3 \times 10^{-1}$, $6 \times 10^2$, $3.6 \times 10^7$ | $1 \times 10^{-12}$ |
| Coulomb force | $10^3$ | $3 \times 10^{-1}$ | $1 \times 10^{-12}$ |

The speed of gravitational force can be easily measured or tested whether it is equal to c or larger than c within spacecraft as $L=3.6 \times 10^7$. For example, as $L=1 \times 10^7$m, if the time duration is $t_2-t_1=1 \times 10^{-2}$ seconds, a speed of $c_g=\sim 3.3c$ can be measured. And, the time duration of $1 \times 10^{-2}$ seconds can be easily measured. (The measured time may be affected by the effect of the general relativity. But, the atomic clock at the height $3.6 \times 10^7$m above the Earth is only faster $g\Delta h/c^2=3.92 \times 10^{-9}$ than that on the Earth,[33] it is too little to affect the measured result.)

The speed of gravitational force certainly can be measured on the surface of Earth as $L=6 \times 10^2$m, because the variation of time on the level of $10^{-6}$ seconds reflecting to the gravity has been detected in Fig.7.3 of Ref.[27]. The speed of gravitational force can be measured in laboratory as $L=3 \times 10^{-1}$m, because the time can be measured on the level of $1 \times 10^{-18}$ seconds as shown in the Appendix 1.

The gravitational universal constant and acceleration are local and timing variation. But, in the experiments in Refs.[25-31], these factors to affect the measured results of the gravitational universal constant and acceleration were well processed. In this design, it is that the variations of the time relative to the variation of the gravitational force are important. The precision for this measurement is only determined with the precision of the measured variations of time. The gravitational force or acceleration under local and timing variations or from different sources does not affect the measured speed of gravitational force. For example, the speed of gravitational force measured from the Sun or from a mass with 500kg is same. It is shown in the Appendix that the variation of time relative to the variations of gravitational force at the level of $10^{-15}$ (or $10^{-18}$) seconds can be measured. Therefore, if the speed of gravitational force is c, correspondent to the distance $3 \times 10^{-1}$ and $6 \times 10^2$ m, it can be very easily measured. As measured in a spacecraft at a height of $3.6 \times 10^7$m, the precision of time is determined by the GPS which is usually $14 \times 10^{-12}$ seconds, it also can be measured easily.

There are two isolated ways for studying the interaction and propagation of the (gravitational and electromagnetic) fields.

The first way is based on the observations of the motion of the planets. The physicists, whether who argued the speed of gravity is c or is faster than c, have to consider these observations: 1) the orbits of the planets in the solar system are stable, it means 2) nonexistence of gravity aberration, the gravitational force is directed to the instantaneous positions of the gravitating bodies, 3) the motion of the Earth reflecting to the motion of the Sun need be instantaneous or much faster than light, and 4) the orbits of the artificial spacecraft (satellite) can be varied without gravity delaying, 5) as the spacecraft is separated or recombined in space, no effect that is not accordant with the Newtonian law of gravitational force has been observed.

In Einstein's general relativity, it is postulated that the speed of gravity is c. But, within this postulation, it is difficult to explain the above observations. Therefore, from the observations 1)-3), Flandern[18] concluded that the speed of gravity is much faster than light. But, it is questioned for that his concluding is based on the assumption of central force and point particle in Newtonian theory of gravity.[20-22] And, compered with the electromagnetic field, it is concluded that, as the speed of gravity is c, the gravitational (and Coulomb) force can be directed to the instantaneous positions of the gravitating bodies and make the orbits of these bodies stable.[20-22] But, the instantaneous direction substantially implies that the speed of gravitational force is instantaneous. For example, during a solar eclipse, the largest gravitational force of both the Sun and Moon acting on the Earth emerges only at the time as the Sun, Moon and Earth are on a straight line Sun-Moon-Earth connecting the Sun, Moon and Earth. Therefore, measured at any point on the line Sun-Moon-Earth, the largest force only emerges at a same time, the measured speed of this largest force is instantaneous. And, in Ref.[7] it was assumed that that the speed of gravity could be observed from that the Sun were to explode into two pieces. This assumption could be observed. In principle, that a planet is to explode in to two pieces is equal to that the Sun and Moon are moving from the positions they are on a straight line Sun-Moon-Earth to the positions they are not on the straight line during a solar eclipse. But, during a solar eclipse, no anomalous variation of orbit of the Earth has been observed as the perihelion advance of Mercury, which is 43″ arc-seconds per 100-year, can be observed. And, as the spacecraft separated and combined in the height of $3 \times 10^7$m above the surface of the Earth, a time of 0.1 seconds would be needed for the gravitational force from the Earth reaching the spacecraft. In this time, the force $GMm/R^2$ could not act on the motion of this spacecraft, $GMm/R^2=mv^2/R$ would be violated, an acceleration of the spacecraft from $mv^2/R$ on the level of *g* (the gravitational acceleration) should be observed in the time of 0.1 seconds. However, no acceleration different from that for the normal motion of the spacecraft has been observed while an anomalous acceleration of $10^{-8}$*g* from other factor on the spacecraft can be observed.[34] It seems that these observations could imply that the speed of gravitational force is instantaneous or much larger than c.

Therefore, it is difficult to explain the observations with the assumption that the speed of gravity is c. The indirect way for knowing the speed of gravity was tried by comparing the gravitational field with the electromagnetic field.[20-22] Intuitively, the interaction and propagation of gravitational field can be understood by this comparing for that a series of laws for the two fields are analogous. (It is noted that the speed of Coulomb force has never been measured although the speed of the electromagnetic waves or radiations has been well measured.) But, this comparing in Refs.[20-22] is ambiguous. In fact, the speed of gravitational force need be compared with the speed of Coulomb force. But, in Refs.[20-22], the speed of gravitational force is compared with the speed of electromagnetic waves or radiations. In this paper, it is explicated that the speed of gravitational force only can be compared with the speed of Coulomb force while the speed of gravitational waves/radiations can only be compared with the speed of electromagnetic waves/radiations.

The second way is the study for only the speed of interaction electromagnetic field. The equations that are same as the Eqs. (10) and (12) were concluded for the electromagnetic field, and it was presented that Eq.(13) could indicated that the speed of the bound (interaction) electromagnetic field is instantaneous.[35] Four experiments[36-39] were performed for measuring the speed of bound electromagnetic field, and in three of them it was measured that the bound electromagnetic fields highly exceed the speed of light or larger than c.[37-39] (Although it was measured that the speed of Coulomb field potential is c,[36] it was pointed out that this result is unreliable.[37])

The speeds for the interaction force/fields obtained from two isolated ways can be proven by each other. It could indicate that the speeds of the interaction force and interaction fields (bound fields) of both gravitational and electromagnetic fields are larger than c. But, no evidence obtained from the two ways could imply that the speed of gravitational force is equal to the speed of Coulomb force and that the speed of gravitational waves/radiation is c.

The design in this paper does not imply that the speed of the interaction (gravitational and electromagnetic) force is larger than c or is c. It only can prove that the speeds of the gravitational and Coulomb force can be tested under the condition that they are larger than c. Because the experiments and observations could show that this speed could be larger than c,[18,37-39] if one wants to confirm whether this speed is c or is larger than c, he/she has to have a method that can measure the speed which is larger than c. This design just meets such a requirement.

Only a finite speed can be measured with this design for that the distances listed in Table 1 cannot be largely increased and the measurable time duration is limited to larger than $10^{-15}$(or $10^{-18}$) seconds. Thus, a speed that is larger than $10^{11}c$ (or $10^{14}c$) cannot be measured or proven with this design. In the experiments[37-39] the precision of time for the speeds of the bound fields is on the level of $\geq 10^{-10}$ seconds. While in the design in this paper, the precision of the time for the speed of interaction force is on the level of $\leq 10^{-12}$ seconds which can measure a larger speed.

**Method:** Measurement of the variation of the time relative to the variation of gravitational force

For precisely measuring the Newtonian gravitational universal constant and gravitational acceleration with atom interferometer, several experiments[25-31] within the design like that shown in Fig. 1a were performed. In these measurements, the motion of the atom acted by a mass with almost 500kg has been measured.

In the atom interferometer, a two level atom with ground state $|a\rangle$ and excited state $|b\rangle$ is initially found in state $|a\rangle$ with energy zero, a local oscillator (LO) provides a frequency $\omega_{LO}$ closes to the resonance frequency $\omega_{ab}$ of the atomic transition $|a\rangle \rightarrow |b\rangle$. An interferometer is built from the interference between two oscillators. Let the local frequency $\omega_{LO}$ accurately match the atomic resonance $\omega_{ab}$, with many atoms instead of one, and lock the LO frequency onto the atomic resonance, an atomic clock is built.[25] In a Raman interferometer, the phase shift from the affection of the mass is given by $\Delta\Phi=k(g_1-g_2)T^2$,[25-30] where k is the Raman laser waves number, g is the gravitational acceleration, T is the time spacing for the pulse. $g_1=g+a_1$, $g_2=g-a_2$. $a_1$ and $a_2$ are from the mass and $a=GM/R^2$, R is varied as the mass is moved.

Typically, the time pacing T is $1.5\times10^{-1}$ seconds. In time T, the measured $\Delta\Phi$ is $3.6\times10^6$ rad. It means that a variation of 1 rad relative to $4.17\times10^{-8}$ seconds can be measured. And, the amplitude of $5\times10^{-3}$ rad can be detected, it means that the variation of $5\times10^{-3}$ rad relative to the time of $4.17\times10^{-11}$ seconds can be measured. Therefore, the experiments[25-31] have shown that a time of $4.17\times10^{-11}$ seconds relative to the variation of the gravitational acceleration *g* and a can be measured.

In these measurements,[25-30] the variations of the atoms from $a_1$ and $a_2$ in two interferometers are measured respectively. And, a line of the variation of $\Delta\Phi$ relative to a time on the level of $10^{-6}$ seconds was measured in figure 7.3 of Ref.[27]. The time relative to the variation of the gravitational force can be measured from comparing the frequency and phase of the atom interferometer with that of a reference. The reference frequency must be of the order of the laser frequency and can be either an atomic transition or another laser. The signal between the interferometer and the reference laser of either 6.567 GHz (cooling and repumping) or around 4.834 GHz (Raman pulses) is detected by a fast photodiode.( Ref.[25], chapters 4.1.5 and 4.3.2) 6.567 GHz or 4.834 GHz is correspondent to a time on the level of $10^{-15}$ seconds. It means that a time on the level of $10^{-15}$ seconds relative to the variation of the gravitational force can be measured by comparing the frequency and phase of the atom interferometer with that of a reference.

With optical atomic clock, difference in gravitational potential for a height of 17cm was measured.[40] It means that: a) The speed of gravitational force can be measured with optical clock. In this case, with the design in Fig.2, the time $\Delta t$ can be measured at the level of $\sim 10^{-18}$ seconds for the optical clock can measure a time at this precision and accuracy.[41] b) The gravitational potential effect on the optical clock need be considered. This effect is $g\Delta h/c^2$.[23] As the measured time for the speed of gravitational force is $\Delta t \geq 10^{-12}$ seconds, this effect at height about $10^2$m is negligible for that it is much less than $10^{-12}$ seconds. While this time is $\Delta t \geq 10^{-18}$ seconds, it is less than the effect $g\Delta h/c^2$ at a height less than $3\times10^{-1}$m. Thus, the gravitational effect at different height has to be considered. And, the variation of the time relative to the variation of gravitational acceleration caused by the moving mass in Fig.1a can be treated as a noise for an optical atomic clock which also can be measured at the precession and accuracy of $10^{-18}$ seconds.[41]

**Note:** There are some mistakes in the Appendix 2 of arXiv:1108.3761v4.

**Appendix A**

# Method to measure the direction of gravitational force and Comments on "Observational evidences for the speed of the gravity based on the Earth tide"

**Abstract**: The direction of gravitational force could be measured by comparing the measured time of the zero tidal force with the measured time of zero tidal angle. And, it is presented that the results in "Observational evidences for the speed of the gravity based on the Earth tide" cannot be obtained from normal process. Tang and coworkers' results are faked.

The direction of gravitational force is a general problem of modern physics. It is a key to understand the orbit of planet and the limit of the speed.[1-3] Tang[4] and coworkers claimed that they observed the direction of the gravitational force by measuring the tidal force. But, calculations show that the variations of the tidal force around zero tidal force in ±506s are less than |-2.55| μGals or |2.63| μGals, and it is less than |0.15| μGals in a time less than 30s. It could show that Tang[4] and coworkers' result are questioned for that the precision of the PET gravimeter that they used is designed to be ±1μGal.[5] Here, from the calculations, a method to measure the direction of the gravitational force is presented.

### 1. How to measure the direction of the gravitational force through the tidal force

Observed at a point O on the surface of the Earth, we can measure a line of a tidal force of the Sun as shown in Fig. A. The tidal force is:

$$F = \frac{GM_S}{(R+r)^2} - \frac{GM_E}{r^2} \qquad (a)$$

where $M_S$ and $M_E$ are the mass of the Sun and Moon respectively, R is the distance between the Sun and point O, r is the radius of the Earth as shown in Fig. B.

With the rotation of the Earth, the tidal force of the Sun is varied as:

$$F = \frac{GM_S r}{R^3}(3cos^2\theta - 1) \qquad (b)$$

At the time $t_\theta$ as the angle θ=54.73⁰, the tidal force is zero. Here, the angle θ=54.73⁰ is called the zero tidal angle. The time $t_z$ of the zero tidal force can be measured as shown in Fig. A, and the time $t_\theta$ can be measured as shown in Fig.B. If $t_\theta = t_z$, it indicates that the direction of gravitational force is directed to the instantaneous positions of the Sun; contrarily, if $t_\theta > t_z$ (or $t_\theta - t_z$=500s), it indicates that the gravitational force is directed to the apparent position of the Sun.

The angle θ is varied with $\theta = \omega dt$, where ω is the speed of the rotation of the Earth. In 506s, ωdt=2.108⁰. Relative to θ=54.73⁰, at $t_\theta$ ±506s, from Eq. (b), it is calculated that the tidal forces are -2.55μGals and 2.63μGals respectively (as θ=0, $\frac{GM_S r}{R^3}(3cos^2\theta - 1) = 50.6 Gals$). Because of the latitudes of the Earth, there is an angle α analogous to angle θ, which is determined by the direction of OS in Fig. B along the NS poles of the Earth. The tidal force need be again revised with $\frac{1}{2}(3cos^2\alpha - 1)$. Therefore, at $t_\theta$ ±506s, the tidal forces are less than |-2.55|μGals and |2.63|μGals. This tidal force cannot be measured with the general gravimeter which only can measure the tidal force that is larger than |±1|μGal. For example, as the calculated value of the tidal force is zero, the measured one could be 1μGal; while the calculated one is 2.7μGals, the measured one could be 1.7μGals. Therefore, the measured results cannot be distinguished from each other. Because it is difficult to detect the gravitational force at the level of ±1μGals with the general gravimeter,[5] atom gravimeter or atomic clock is needed to put this method into practice.

The time $t_\theta$ can be measured by measuring the zero tidal angle θ=54.73⁰ in Fig. B. To measure angles θ and α, first, we need measure the position of the Sun relative to the observer which can determine the direction of line OS. Second, we need measure the direction of line OC. Third, we need measure the length of the lines OS and OC. It is difficult to accurately measure these values for: a) distance between the Sun and the observer. For example, in 2012, it is redefined the distance between the Sun and Earth.[6] b) the gravity anomaly of the Earth. It is about 0.5%, it is correspondent to that a variation of about 20km of the radius of the Earth and variations of about 0.5% of the tidal force.[7] c) The parameters for long period earth tidal are difficult to be obtained with high accurate.[8]

It is noted that the line of a tidal force can be calculated within a model by only the time and the longitude and latitude of the observation point. This model was programed into software.[9,10] But, because the Earth is not homogeneous and perfect sphere, a set of parameters is needed for this model (software). The result calculated within the model (software) is not absolutely accurate. For example, the largest difference between the theoretical values calculated with the Tsoft and MT80W is 2.6μGals. [10] Therefore, it cannot be used to know the direction of the gravitational force.

2. **Comments on "Observational evidences for the speed of the gravity based on the Earth tide"**

Tang[1] and coworkers claimed that they obtained the observational evidences for the speed of the gravity. They claimed that gravitational force is directed to the apparent position of the Sun.

The works that they claimed they did are not true. According to their claim, first, they "constructed a quasi-solar tide curve" from an original observed data. Second, according to the Equation (1) in their paper, they calculated the second line of this force from the practical Newtonian formula. And, according to classic Newtonian formula, they calculated the third line of this force. Third, they compared the times of these lines at the "zero-tide point" with each other. In principle, the direction of gravitational force cannot be measured through this process. The tidal force of a certain position and time of the Sun need be measured for a calculated line of the tidal force. But, as shown in the above, if such a measurement was performed, the direction of the gravitational force had been measured. Thus, the calculated lines are not needed. Conversely, if they did not perform such a measurement, they cannot obtain the calculated lines as the claimed.

The data in Ref.[4] was not obtained from normal calculation. First, as calculated above, in 506s around the zero tidal force point, the variation of this force is almost 2.5μGals. Thus, the variations of the solar tidal force cannot be measured with their method. Second, in Figures, 2 and 3, they claimed that, they obtained the time differences of 24s and 18.2s between the measured and calculated zero tide points. But, in less than 30s, according to Eq. (b), the variations of tidal force around the zero tidal force point is less than 0.15 μGals. It seems that they did not calculate the lines of the solar tidal force from the equation (1) in their paper or from classic Newtonian formula. Third, as point above, the calculated lines of the tidal force from a tidal model or software are invalid. These calculated lines must be based on the measured local radius of the Earth, the local distance of the Sun relative to the observer and the related data to determine angle θ in Fig. B. But, they did not have such a measurement. And, they did not discuss the precision of the tidal model with the software for their calculation. If they did have a calculated result from a measurement or a model (software), they should have found that the difference of the tidal force between the observed curve and the calculated curves in Figures 2 and 3 cannot be distinguished from each other.[5] Because they did not do the above works, the data in Figures 2 and 3 have nothing to do with the observed original data in Figure 1 in their paper or other measurement and calculation. These data was not obtained from a normal process.

They did not normally process the solar tidal force. They did not obtain a line of a solar tidal force from the original observed data in Fig. 1. First, they did not provide such a line. Second, it is known that there are 4-5 kinks of affections, including the atmosphere, polar motion, instrument drift, long term gravity change, rainfall, earth and ocean tides, etic., on the original measured result. [11] The totality of these affections can arrive at almost 30μGals. But, they did not filtered out such affections. Third, especially, the earth solid tides are one of the affecting factors. It need be filtered out to obtain an accurate solar tidal force. But, they used it as the main factor for the direction of gravitational force. It is clear, they did not obtain an accurate and precise line of the solar tidal force. They only obtained "a quasi-solar tide curve". In their work, the "quasi-solar tide curve" can be 69.11μGals.[12] While calculated from Eq.(b), the solar tidal force only can be less than 51μGals.

In summary, Tang and coworkers claimed that the data they used are recorded by the PET gravimeters.[4]

The precision designed for the PET gravimeters is ±1μGal.[5] The data in the Figures 2 and 3 in Tang's and coworkers' paper[4] are less than |±1.5|μGals. If 1μGal is reduced from the red line in the Figures 2 and 3, from Eq.(b), it is calculated that the zero point of the red line shall be positioned almost at the middle between the green and blue lines. Therefore, the PET gravimeter cannot be valid for their purpose. In fact, their work [4] has nothing to do with the precision of the measurement. They did not obtain an accurate and precise line of the solar tidal force. Their work has nothing to do with the measurement recorded by the PET gravimeters or the result calculated within an Earth model or software of tide. The Figs. 2 and 3 and the result in their paper are faked.

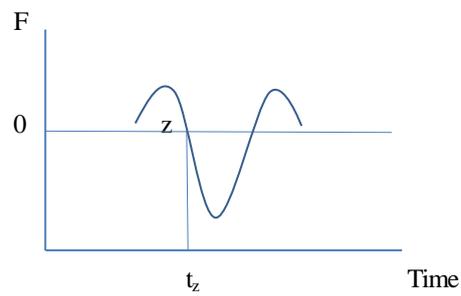

**Fig. A. The measured line of the tidal force of the Sun.** At the point z the tidal force is zero. It is relative to the time $t_z$.

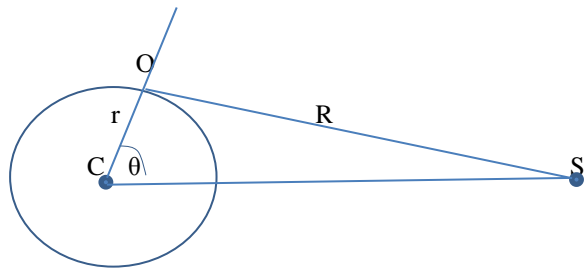

**Fig. B. The variations of the tidal force.** The observer is at point O. C is the center point of the Earth. The Sun is at point S. r is the radius of the Earth, R is the distance between the Sun and observer. θ is varied with the rotation of the Earth. As $θ=54.73^0$, the tidal force is zero. And a time $t_θ$ relative to $θ=54.73^0$ can be measured.

**Appendix B**

# The speed of Gravity: An Observation on Satellite Motions

**Abstract**

The radius of orbit of the geosynchronous satellite can be observed at the precision of less than 8cm. And, a force about $\sim 10^{-9}$m/s$^2$ can make the orbit of satellite shifted. Here, the gravitational forces of the Sun acting on the satellite from the present and retarded positions are calculated respectively, assuming that the retarded position is determined with that the speed of the gravitational force is equal to the speed of light. It is shown that the difference of the force between the present and retarded positions of the Sun acting on a geosynchronous satellite can be larger than $1 \times 10^{-7}$m/s$^2$. And, the difference of the radius of the orbit of the satellite perturbed by the gravitational force of the Sun from the present and retarded positions in 3000s can be larger than 8.2m. It indicates that the gravitational force of the Sun acting on the satellite is from the present position of the Sun and that the speed of the gravitational force is much larger than the speed of light in a vacuum.

The speed of gravity is an important physical constant. But, the gravitational waves have not been detected, we have had only an indirect evidence for this speed.[1-2] Because the gravitational force can be observed through the moving of the gravitating planets, now it is a main line to observe this speed. It was presented that the speed of gravitational force need be much larger than the speed of light c to let the orbits of planets stable.[3-4] And, it was argued that, although the speed of gravitational force is c, the gravitational force could be extrapolated from the retarded position of the Sun to the instantaneous position of the Earth.[5-7]

The observation from the orbits of planets in the solar system is based on one unique evidence: These orbits are stable.[4] Thus, the conclusion from this line is easy to be disputed.[5-7] The orbit of a satellite is based on a series of data calculated from the Newtonian theory of gravity. These data can clearly and certainly show the certain point where a gravitational force of the Sun acting on the satellite is from. Therefore, the observation based on the orbit of a satellite is certain and clear and beyond dispute.

Within the Newtonian theory of gravity, the acceleration at the precision of $\sim 10^{-10}$m/s$^2$ on the satellite can be measured;[8] The actual and predicted data of the radius of the GPS satellite at the precision of less than 3cm usually can be provided by the GPS office.[9-10] A force perturbing the orbit can be calculated individually.[11-12] Here, the gravitational forces of the Sun from the present and retarded positions perturbing the orbit of geosynchronous satellite are calculated respectively, assuming that the retarded position of the Sun is determined with that the speed of gravitational force is c. By comparing the calculated gravitational force of the Sun acting on the satellite from the present position of the Sun with that from the retarded position of the Sun, it is known that the gravitational force of the Sun acting on the satellite is from the present position and the speed of gravitational force is larger than c.

The speed of gravitational force was distinguished from that of gravitational waves.[2] And, an analogous distinction between the speed of interaction (force) electromagnetic field and that of electromagnetic waves was worked out by two groups.[13-17] The observation in this paper can be explained and understood with the way presented in Ref.2 and the results in Refs.13-17.

For the orbit of a geosynchronous satellite,[18] e=0, i=0, where e and i are the eccentricity and inclination of the orbit respectively. Thus, this is a circular orbit which is in the equatorial plane of the Earth. At the same time, for the convenience, I first let the line connecting the Sun and the center of the Earth be in the same plane. (I shall discuss the status that this line is not in the same plane in the Supplementary Information)

This orbit perturbed by the gravitational force of the Sun from the present and retarded positions can be shown in Fig.1.

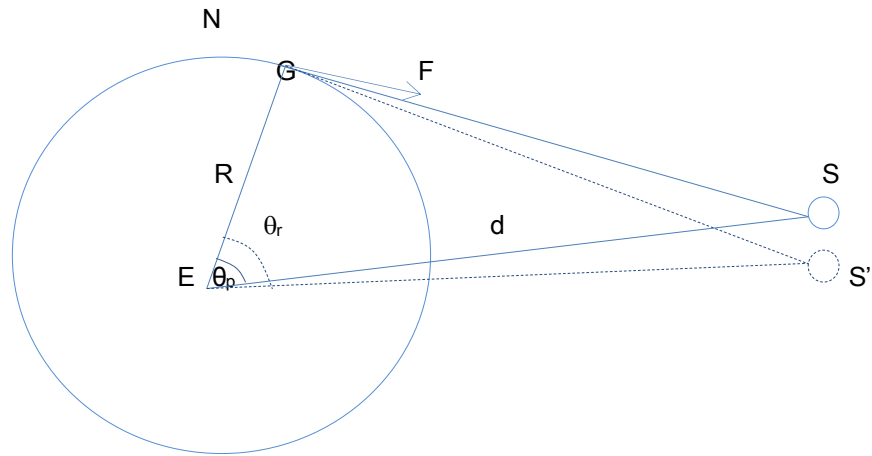

**Fig. 1. The gravitational force of the Sun from the present and retarded on the satellite.** The orbit of the satellite is circular. The satellite is at point G. The S and S' are the present and retarded positions of the Sun respectively. S' is determined with t'=t-d/c, where c is the speed of light in a vacuum. p and r denote the present and retarded positions of the Sun respectively. E is the center of the Earth. F is the gravitational force of the Sun acting on the satellite and it is vertical to R. N is the north.

Under the conditions of Fig.1, the gravitational force, F, of the Sun acting on the satellite at point G vertical to the line R is varied as[19]

$$F = \frac{3}{2}\frac{GMR}{d^3}\sin 2\theta, \tag{1}$$

As shown in Fig.1, assuming that the Earth is stationary and the angular speeds of the Sun and the satellite are respectively $\omega$ and $\omega_1$, the Sun is moving from east to west while the satellite is from west to east, therefore, $\theta=\theta_0-(\omega_1-\omega)dt$. For a geosynchronous satellite, $R=4.2\times10^4$km, $\omega_1=0$. On average, $\Delta t=d/c=500$s, $\omega=\frac{-1}{240}\times1^0$/s, and $\frac{3}{2}GMR/d^3=2.5\times10^{-6}$m/s$^2$, $\Delta\theta=(\omega_1-\omega)\Delta t=2.08^0$. And, $\theta_r=\theta_p+\omega\Delta t$, (i.e., $\theta_r=\theta_p+2.08^0$). From $\theta_0=70^0$ to $59.6^0$ at the step of $\Delta\theta=(\omega_1-\omega)\Delta t=2.08^0$, according to Eq.(1), and $\theta_r=\theta_p+2.08^0$, the gravitational force F of the Sun from the present position and from the retarded position acting on the satellite is shown in Table 1.

**Table 1. The difference between $F_p$ and $F_r$ ($10^{-6}$m/s$^2$)**

| $\theta_p$ | $70^0$ | $67.92^0$ | $65.84^0$ | $63.76^0$ | $61.68^0$ | $59.60^0$ |
|---|---|---|---|---|---|---|
| $F_p$ | 1.61 | 1.74 | 1.87 | 1.98 | 2.09 | 2.18 |
| $\theta_r$ | $72.08^0$ | $70^0$ | $67.92^0$ | $65.84^0$ | $63.76^0$ | $61.68^0$ |
| $F_r$ | 1.46 | 1.61 | 1.74 | 1.87 | 1.98 | 2.09 |
| DF | 0.15 | 0.13 | 0.13 | 0.12 | 0.11 | 0.09 |

Note: p and r denote the present and retarded positions of the Sun respectively. DF=$F_p$-$F_r$.

According to $\Delta t=d/c=500$s and $\Delta\theta=(\omega_1-\omega)\Delta t=2.08^0$, from Table 1, the relationship between the force and time can be obtained as shown in Fig.2.

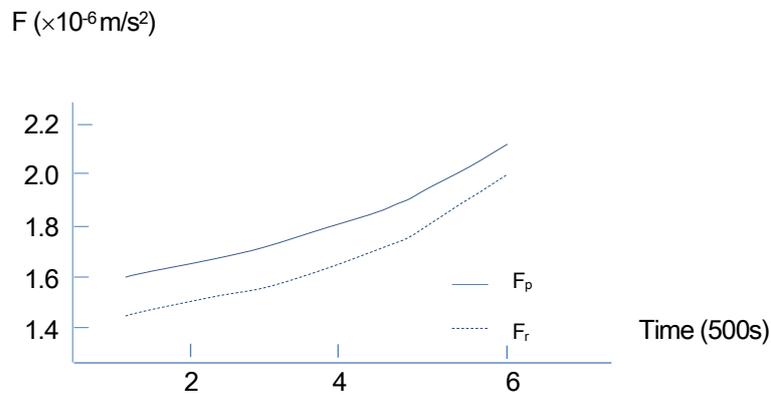

**Fig.2. The difference of the gravitational force between the present and retarded positions of the Sun acting on the satellite.**

The perturbed orbit of a satellite is studied with Lagrange's planetary equations. In these equations, The semi-major-axis of a perturbed orbit is varied as:[11-12]

$$\frac{da}{dt}=\frac{2}{n}[gesinf+F(1+ecosf)], \qquad (2)$$

Where n is the angular speed of the satellite orbiting the Earth, e is the eccentricity, f is the true anomaly, g and F are the gravitational force of the Sun acting on the satellite along the radius and along the direction of the satellite moving respectively.

For the orbit of a geosynchronous satellite as shown in Fig. 1, e=0, a=R, n=ω. Eq.(2) becomes: $\frac{dR}{dt} = \frac{2}{\omega} F$. Thus, we have:

$$\Delta R = \frac{2}{\omega} \bar{F} \int_0^{\Delta t} dt, \qquad (3)$$

Where, $\bar{F}$ is the average value of the gravitational force of the Sun acting on the satellite as shown in Fig.1.

In $\Delta t = d/c = 500s$, there is $\Delta\theta = (\omega_1 - \omega)\Delta t = 2.08^0$. And, in this time, the average of the force in Eq.(3) is

$$\bar{F} = \frac{1}{\frac{2.08^0}{\omega}} \frac{3}{2} \frac{GMR}{d^3} \int_0^{\frac{2.08^0}{\omega}} \sin 2(\theta_0 - \omega t)\, dt, \qquad (4)$$

From Eqs.(3) and (4), the radius R of the orbit of the satellite perturbed by the gravitational force of the Sun from the present and retarded positions respectively is shown in Table 2.

**Table 2. The difference between ΔR$_p$ and ΔR$_r$ (m)**

| $\theta_p$ | 70.00⁰—67.92⁰ | 67.92⁰—65.84⁰ | 65.84⁰—63.76⁰ | 63.76⁰—61.68⁰ | 61.68⁰—59.60⁰ |
|---|---|---|---|---|---|
| ΔR$_p$ | 23.11 | 24.83 | 26.48 | 27.99 | 29.37 |
| $\theta_r$ | 72.08⁰—70.00⁰ | 70.00⁰—67.92⁰ | 67.92⁰—65.84⁰ | 65.84⁰—63.76⁰ | 63.76⁰—61.68⁰ |
| ΔR$_r$ | 21.12 | 23.11 | 24.83 | 26.48 | 27.99 |
| DR | 1.99 | 1.72 | 1.65 | 1.51 | 1.38 |

Note: p and r denote the present and retarded positions of the Sun respectively. DR=ΔR$_p$- ΔR$_r$

The radius of the satellite perturbed by the gravitational force can be summed up.[12] From Table 2, we have $\sum\Delta R_p = 131.78m$, $\sum\Delta R_r = 123.53m$. Therefore, in 3000s, from $\theta = 70^0$ to $59.60^0$, the difference of the perturbed radius of the orbit is increased from 1.99 m to 8.25m as shown in Fig.3.

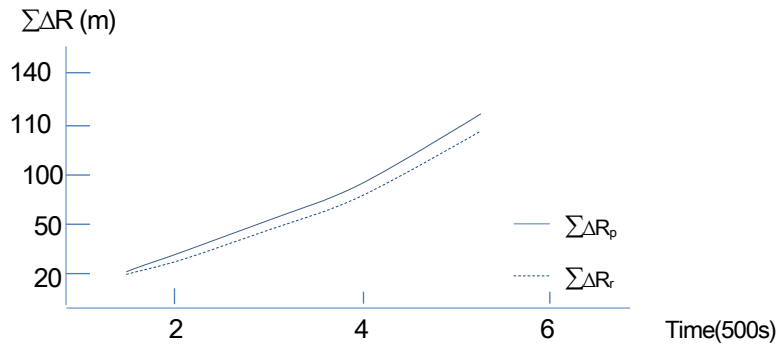

**Fig.3. The difference of the radius of the orbit perturbed by the force of the Sun from the present and retarded positions respectively.** In Eq.(1), the gravitational force of the Sun acting on the satellite is varied periodically with $\sin 2\theta$. Hence, in some time, such as from $\theta=150^0$ to $180^0$, $F_p<F_r$ and $\Sigma\Delta R_p<\Sigma\Delta R_r$.

It is noted that the calculated result is determined by both the values R and the angle speed of the satellite, $\omega_1$. The differences of the perturbed force and radius are larger as the R is larger. And, they are less as $\omega_1$ is larger. For example, as the radius is R=100km, $\frac{3}{2}GMR/d^3=0.375\times10^{-6}$m/s$^2$, the largest difference of the forces is less than $2\times10^{-8}$m/s$^2$. And, as $\omega_1$ is less, the period of $\sin 2[\theta_0-(\omega_1-\omega)]$ become longer. It results in that $\Sigma\Delta R$ is larger for the factor of $2/\omega$ in Eq.(3) or n'/n in (1s) and for the same angle $\Delta\theta$ need be larger time to orbit. Therefore, the results in Tables 1 and 2 and in Figs.2 and 3 are less than that for an actual geosynchronous satellite for it there are $R>4.2\times10^4$km.

Kozai computed the lunisolar perturbations in satellite motions.[20] In his work, it is shown that the Eqs.(3) and (4) can accurately describe the orbit of geosynchronous satellite. Although the result is affected by the angle between the equatorial plane and the line connecting the Sun and the center of the Earth, its effect cannot change the result in this observation. (Please see the Supplementary Information.)

A gravitational force of $\sim 10^{-7}$m/s$^2$ has a big perturbation on the orbit of a satellite. For example, as a satellite is on the altitude less than 1000km, the gravitational force of the Sun on it is $\sim 10^{-7}$m/s$^2$, and the pressure of the radiation of the Sun on a satellite is $\sim 10^{-7}$m/s$^2$. And, the forces of the solid and ocean tides are $\sim 10^{-9}$m/s$^2$. These forces can make the orbit shifted and can be observed and calculated accurately.[9,10,12] The real semi-major-axis of orbit of a GPS satellite can be obtained online from the office of GPS.[9] From the real initial orbit, in next 12 hours, the orbit with the precision of less than 3 cm in the step of 15minutes can be calculated with IGS ephemerides and be observed.[9-10] An observation of geosynchronous satellite shows that, for 2-h orbital predication, the error of predicted radius is about 40.4 cm, and the error of the radius evaluated with post-processed data is 7.6 cm.[21] Therefore, it is clear that a difference of the radius of 8.25m and a difference of the force larger than $1\times10^{-7}$m/s$^2$ can be easily measured. This observation certainly and clearly shows that the gravitational force of the Sun acting on the satellite is not from the retarded position.

Within the observable precision that the radius of a satellite is less than 8cm, from this observation we cannot conclude that the speed of gravitational force is infinite or how large it is. And, from Eq.(1) and (2) and the condition in Table 2, to arrive at this precision, the speed of the gravitational force only need be larger than 100c. According to Ref.4, if this speed is less than $2\times10^9$c, the orbits of the Earth shall shift. It is clear, if the orbit of the Earth was varied 8cm in 500 seconds, it should shift in a short time. Therefore, to know the speed of gravitational force in higher precision and accurate, it need be measured with the design in Ref.2 or other way.

In Fig.1, assuming that the Earth is stationary, the geosynchronous satellite also is stationary. Therefore, it is the motion of the Sun/Moon that perturbs the motion of the satellite. (It is intuitive as the Sun was replaced by the Moon in Fig.1.) As the Sun/Moon begins to move, the satellite shall be moved. There is t'=t-d/c, where, t is the time of the Sun begins moving, t' is the time that the satellite begins to be moved, c is the speed of light in a vacuum, d is the distance between the Sun and the Earth. t'=t-d/c is usually called the retarded time. There are three kinds status for the retarded time. 1) $v_g$=c. this is the conventional interpretation about the retarded time. 2) $v_g$ is infinite, t'-t=0. It means that the respondence of the satellite to the motion of the Sun is instantaneous. And, 3) c<$v_g$<∞, it means that the respondence of the satellite to the motion of the Sun is very fast. For example, as $v_g$=1$\times10^9$c, t'-t=d/$v_g$=5$\times10^{-7}$ seconds.

This result can be measured with the design in in this paper. The gravitational force between the atom and a mass with 500kg can be measured with atom interferometer. The motion of a mass with 500kg can be operated easily. Thus, the respondence of the atom to different motions (such as that with acceleration or uniform velocity, or stopped suddenly) of this mass can be measured.

This method also is suitable for the speed of Coulomb force. The respondence of a stationary charged ball to a moving charged ball can be observed with the same way.

In this respondence, the energy, momentum and communication are transferred from the moving object to the responded one. For example, as the satellite is responded to the motion of the Sun/Moon, the satellite is moved with a speed v. Thus, it is moved with an energy E=$mv^2$/2 and a momentum P=mv, where m is the mass of the satellite. v is determined with $\mathbf{v}=\partial\mathbf{R}/\partial t$, $\mathbf{R}$ is determined with Eq.(3), where $\mathbf{v}$=v(x,y,z), $\mathbf{R}$=R(x,y,z). In Table 2, the difference between $R_p$ and $R_r$ shows that the velocity v of the satellite is responded to the motion of the Sun at its present position.

**Discussions and Conclusions**

The direction of the field of a charge or a mass moving with a speed v was usually studied with the Liénard-Wiechert potential,[6] and there is[22]

$$g = e\frac{1}{\gamma^2(1-\beta\cdot n)^3 R^2}(\mathbf{n}-\boldsymbol{\beta}) + e\frac{1}{(1-\beta\cdot n)^3 Rc}\mathbf{n}\times[(\mathbf{n}-\dot{\boldsymbol{\beta}})\times\boldsymbol{\beta}], \quad (5)$$
$$= \text{interactive term} + \text{radiative term}$$

The radiative term was well used to study the radiation of a charge and it was well known that the speed of the electric radiation is c.[22] But, the interactive term has not been known experimentally.

The interactive term is correspondent to the Coulomb force.[22] The speed of Coulomb force has never been measured.[2,12-17] But, to explain the speed of gravity, the speed of the interactive term was arbitrarily given a speed that is equal to the speed of light.[5-7] It was presented that the explanation in Refs.5-7 is questioned,[2] and, by two groups within different ways, it was measured that the speed of the interactive term is larger than c.[13-17]

Now, a "common view" is current for many of physicists. In this "common view", it is believed that the direction of the electric field of a charge moving uniformly is directed to the present position of the charge while that a charge moving with an acceleration is not so.[6] But, in Eq.(5), the interactive term is independent of the acceleration. In Ref.22, it is clearly shown that the interactive term in Eq.(5) is the same as that of the charge moving uniformly. Thus, the "common view" of the direction of the field of the charge moving with an acceleration is mistake.

It is noted that this "common view" is used as a criteria to assess other works. This is not a normal status. There is no experiment or observation for this "common view". Even this "common view" has not been restricted with a normal scientific paper or book. Therefore, taking such a "common view" for criteria shall result in that the physics is far from the scientific foundation which is based on experiment and observation. To clearer discussing this problem, the speed of gravity is studied with the structure of field in the Appendix C.

In summary, this is the first direct observation of the speed of gravity. And, any valid "common view" or explanation for the speed of gravity must be accordant with experiment or observation. This observation could be accordant with the recent experiments. Besides that it was measured that the speed of the interaction (force) electromagnetic field is larger than c,[13-17] the nonlocality (The speed of spooky action at a distance is larger than $10^4$c.) was measured in quantum entanglement;[23-24] And, it was presented that any finite speed v with $c < v < \infty$ predicting correlations can be exploited for faster-than-light communication.[25]

Supplementary Information for

**The speed of gravity: An observation from satellite motions**

## The orbit of satellite perturbed by the Sun

Y. Kozai computed the lunisolar perturbations in satellite motions.[20] Taking the geocentric rectangular coordinate, for a high altitude, small inclination and almost circular orbit, the radius of the orbit perturbed by the gravitational force of the Sun in a short time is

$$\frac{dr}{a} = -\left(\frac{n\prime}{n}\right)^2 \left(\frac{a\prime}{r\prime}\right)^3 \beta \left[1 - \frac{3}{2}cos^2\delta + cos^2\delta cos2(L + \Omega - \alpha)\right], \qquad (1s)$$

Where a, r, n and a', r', n' are the semi-major axis, radius and angular speed of the satellite and Sun respectively. $\delta$ is the declination, $\alpha$ is the right ascension, L is the argument of latitude, $\Omega$ is the longitude of ascending node. For convention, from Fig.1s, under the condition that r' and the orbital plane is in the equatorial plane, we have $\delta=0$, $\Omega=0$. Therefore, Eq.(1s) becomes

$$\frac{dr}{a} = -\left(\frac{n\prime}{n}\right)^2 \left(\frac{a\prime}{r\prime}\right)^3 \beta \left[-\frac{1}{2} + cos2(L - \alpha)\right], \qquad (2s)$$

As e=0, $(n'/n)^2$ and $(a'/r')^3$ are constant in a short time, $\beta$ also is a constant, they can be written as k. As shown in Fig.1s, if i=0, the orbit and the equator are in the same plane. The angle $\theta$ between r and r' in Fig.1s is just that between R and d as shown in Fig.1. From Fig. 1s, we know L= L=υ+f, thus, L is determined with the angular speeds of the satellite $\omega_1$. And, $\alpha$ is the angle that is measured eastward along the equatorial plane from the vernal equinox as shown in Fig. 1s to the hour circle of the point in question. Thus, as i=0 and assuming that the Earth is stationary, the angle speed $\omega$ of the Sun can be expressed with $\alpha$. Therefore, Eq.(2s) can be rewritten as

$$\frac{dr}{a} = -k\left[-\frac{1}{2} + cos2(\omega_1 - \omega)dt\right], \qquad (3s)$$

Because $sin\theta=cos(\theta+\pi/2)$ and the initial angle $\theta_0$ for both Eqs.(3s) and (4) can be selected freely, therefore, Eq.(4) also can be rewritten as

$$\bar{F} = \frac{1}{\Delta t}\frac{3}{2}\frac{GMR}{d^3}\int_0^{\Delta t} cos\ 2(\theta_0 - \omega t)\ dt, \qquad (4s)$$

Noting that, in Eq.(4s), $\omega=\omega_1-\omega$, therefore, Eqs.(3) and (4) is same as Eq.(3s) for describing the radius of the orbit perturbed by the Sun.

δ=0 means that r' is in the equatorial plane. This status emerges in a short time during equinox. In other times, δ≠0. Therefore, in an actual observation, for more accurately considering the perturbation, δ≠0 need be considered. For a geosynchronous satellite, δ is varied from $0^0$ to $23.5^0$ relative to the time from the equinox to the solstice. δ can be known from the ephemeris.

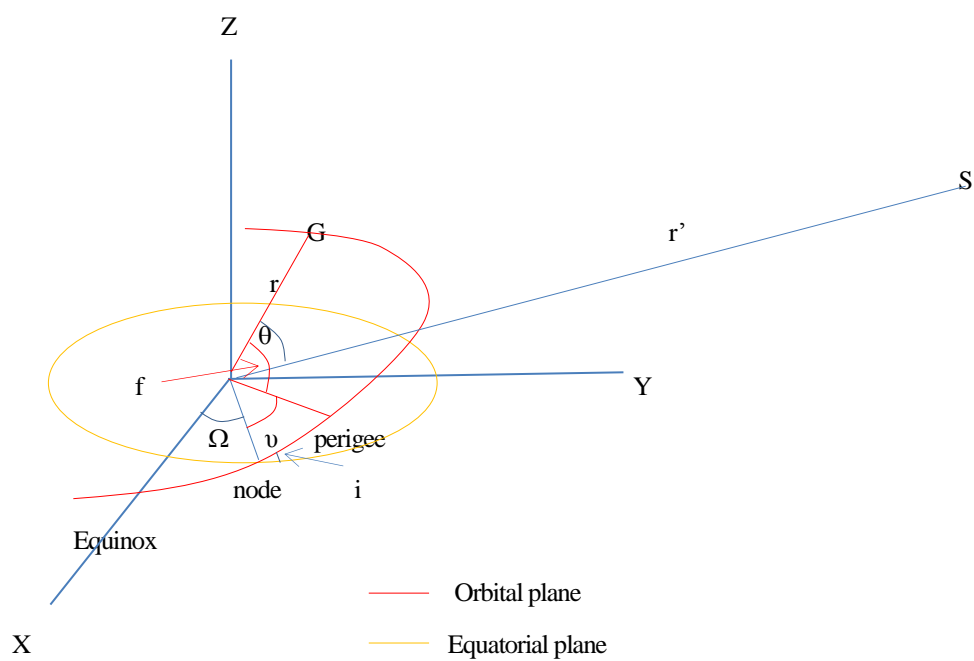

**Fig.1s. The orbit of a satellite.** The Equatorial plane is in the XY plane. The Sun is at S, the satellite is at G. L=υ+f. θ is the angle between r and r'.

Appendix C

## Testing the structure of field and the wavelengths of gravitational waves

**Abstract:** The speed of gravity is studied by comparing the structure of the field of a moving source with the interaction and radiation of the field. From recent experiments and the observation in Appendix B, a line to indirectly test the wavelength of gravitational waves is presented.

The propagation and speed of the interactive field/force of a moving source (charge or mass) can be studied with the structure of field.[1]

The gravitational and electric fields were studied more than 100 years ago, the systematic theories were established for it and they were well used in many areas. But, the structure of the field has not been known experimentally. It is known that the field of a moving source consists of near and far fields with an intermediate zone between them. But, we have not known the relationship between the near field and the far field. For example, how is the far field changed from the near field? Do the far field and near field independently emit from the source? If the far field is changed from the near field, then, how does the near field changes to the far field? Is the speed (c) of the far field of a charge suddenly or gradually changed from that (infinite) of the near field?

The near and far fields are defined with the wavelength. If the source dimensions are order of d and the wavelength is $\lambda = 2\pi c/\omega$, and if $d \ll \lambda$, there are three spatial regions for the field [1]:

| | | |
|---|---|---|
| The near (static) field: | $d \ll r \ll \lambda$ | |
| Intermediate (induction) field: | $d \ll r \sim \lambda$ | (c1) |
| Far (radiation) field: | $d \ll \lambda \ll r$ | |

The radiative and static/interactive fields also is studied with the Liénard-Wiechert potential[1, chapter 9], and

$$g = e \frac{1}{\gamma^2 (1-\boldsymbol{\beta}\cdot n)^3 R^2} (\mathbf{n} - \boldsymbol{\beta}) + e \frac{1}{(1-\boldsymbol{\beta}\cdot n)^3 Rc} \mathbf{n} \times [(\mathbf{n} - \boldsymbol{\beta}) \times \dot{\boldsymbol{\beta}}], \qquad (c2)$$
=interactive term + radiative term

The near and far fields are correspondent to the interactive and radiative fields respectively. Therefore, the structure of the field can be studied with the interactive and radiative fields in Eq.(c2).

Now we have had these results about the interactive and radiative fields:

1) It is well known that the speed of the radiative electromagnetic field is c.
2) The first indirect evidence for the speed of gravity was obtained from the measurement of gravitational radiation in the binary pulsar PSR 1913+16[2].
3) It was measured that the speed of interactive electromagnetic is larger than c and there are different speeds relative to different wavelengths [3-5] or is infinite[6].
4) It was observed that the speed of gravitational force is much larger than c.(Appendix B)

It is noted that, in the experiments in Refs. [3-5], the different speeds relative to different wavelengths of the electromagnetic field were measured. It could be an indirect evidence for the structure of the field. It indicates that the speed of the field is not changed suddenly from an infinite speed to c, but is gradually changed. In Eqs.(c1), different wavelengths are relative to different spatial regions from the near field to the far field respectively, these experiments should imply that the speeds and wavelengths are gradually changed from near field to the far field.

The near and far fields are the spatial regions. Thus, the two fields can be identified by measuring the speeds of interactive force at different distances from the source with the design in this paper. If different speeds of the interactive force relative to different distances from the source are measured, the structure of the fields shall be identified.

This method can be combined with R. Smirnov-Rueda and coworkers' method [3-5]. From the two methods, the speeds and wavelengths of the field in different spatial regions shall be known.

Now, the experimental evidence is unsufficient, we have not completely known the structure of the field. There are some of contradicted explanations about the field. For example, it is widely believed that the gravitational field is static which is not in propagation [7]. But, as pointed out in the Appendix B, the interactive terms for both the uniform moving and accelerated sources are same. Here, it is shown that the relationship between the interactive (static) and radiative fields has not been considered by the physicists who believed that the gravitational field is static. Thus, it is stressed that, because of these questions, the thought that the gravitational field is static or a curvature space is invalid.

From the relationship between the structure of field and the interactive and radiative fields in Eq.(5), we can have these conclusions for the structure of the electric and gravitational fields:

1) The wavelengths of the gravitational waves can be indirectly known. If the speeds and wavelengths of electric field in different spatial regions are known, and the speeds of the gravitational force and radiation are known, the wavelengths of gravitational waves in different spatial regions shall be concluded from these evidences.

2) Some of evidences may have been obtained for the structure of the field. The speed of the interactive (force) field may be known with Refs. [3-6] and by the observation in this paper. The speed of the radiative field is well known for electromagnetic radiation and indirectly known for gravitational radiation) [2]. Intuitively, the radiative field is only the behaving of the interactive (static) field in the far regions from the source. The experiments [3-6] indirectly showed that the interactive (static) field is radiative and the speed of the interactive field is gradually changed. But, the experiments for the structure of field are still unsufficient, further experiments are needed to confirm it.

3) It is not right that the gravitational field is static or a curvature space which is not in propagation. Besides the wrong explanation about the interactive (static) field as pointed in Appendix B, there are three problems: First, Eq.(c1) shows that a field is more complicated than that described with Eq.(c2). The near and far fields are correspondent to the interactive (static) and radiative fields respectively. But, the intermediate field does not have a correspondent term in Eq.(c2). Therefore, a gravitational field cannot be explained and studied only with the static (interactive) and radiative fields. Second, it is noted that the gravitational radiation/wave was studied with the weak-field approximation [8,9] and the first indirect evidence for the speed of gravity was obtained from the measurement of gravitational radiation in the binary pulsar PSR 1913+16 [2]. No experiment shows that the interactive (static) field and the radiative field are isolated from each other. Third, it is the most important, no experiment and observation show that a static (interactive) field is not radiative and with a unique speed. Contrarily, the experiments [3-5] indirectly showed that the interactive (static) field is radiative and with different speeds relative to different distance from the source, therefore, it is in propagation.